\documentstyle[prl,aps,multicol,mypsfig2]{revtex}

\begin{document}

\preprint{}

\title{Experimental realization of order-finding with a quantum computer}

\author{Lieven M.K. Vandersypen$^{1,2,*}$, Matthias Steffen$^{1,2}$, Gregory Breyta$^2$, Costantino S. Yannoni$^2$, Richard Cleve$^3$,\\ and Isaac L. Chuang$^2$}

\address{\vspace*{.4ex}
{$^1$ Solid State and Photonics Laboratory, Stanford University, Stanford, CA 94305-4075}\\[.2ex]
{$^2$ IBM Almaden Research Center, San Jose, CA 95120}\\[.2ex]
{$^3$ Dept. of Computer Science, University of Calgary, Calgary, Alberta, Canada T2N 1N4}\\[.2ex]
}

\date{\today \vspace*{-.5ex}}

\maketitle

\begin{abstract}
We report the realization of a nuclear magnetic resonance (NMR) 
quantum computer which combines the quantum Fourier transform 
(QFT) with exponentiated permutations, demonstrating a 
quantum algorithm for order-finding.
This algorithm has the same structure as Shor's algorithm and its
speed-up over classical algorithms scales exponentially.
The implementation uses 
a particularly well-suited five quantum bit molecule and was made possible
by a new state initialization procedure and several quantum control 
techniques.
\end{abstract}

\begin{multicols}{2}

\def\be{\begin{equation}}
\def\ee{\end{equation}}
\newcommand{\ket}[1]{\mbox{$|#1\rangle$}}
\newcommand{\mypsfig}[2]{\psfig{file=#1,#2}}

The quest for the experimental realization of quantum computers has 
culminated in the creation of specific entangled quantum states,
most recently with four quantum bits (qubits) using trapped
ions\cite{Sackett00}, and seven qubits\cite{Knill00} using liquid 
state nuclear magnetic resonance (NMR)\cite{Gershenfeld97,Cory97}, 
and in the successful implementation of Grover's search 
algorithm~\cite{Chuang98a,Jones98a,Vandersypen00} and the
Deutsch-Jozsa algorithm~\cite{Chuang98b,Jones98b,Marx00} on two,
three, and five qubit systems (see~\cite{Jones00} for additional 
references).

However, a key step which remains yet to be taken is a computation
with the structure of Shor's factoring algorithm\cite{Shor94,Ekert96}, 
which appears to be common to all quantum algorithms that achieve
exponential speed-up~\cite{Cleve97}. 
This structure
involves two components: exponentiated unitary operations and the quantum
Fourier transform (QFT).  
Implementing these components is challenging
because they require not just the creation of {\em static}
entangled states~\cite{Sackett00,Knill00}, 
but also precise {\em dynamic} quantum control over the
evolution of multiple entangled qubits, over the course of tens to hundreds of
quantum gates for the smallest meaningful instances of this class of 
algorithms.
The evolution of the states is precisely where NMR quantum computers appear 
to have an exponential advantage over classical computers~\cite{Schack99}.

Here we report the experimental implementation of a 
quantum algorithm for finding the order of 
permutation\cite{Shor94,Ekert96,Kitaev95}; its structure is the same
as for Shor's factoring algorithm and it scales exponentially faster than any
classical algorithm for the problem.  The realization of 
this algorithm was made possible by the synthesis of an unusual molecule 
with five pairwise coupled, easily addressable $^{19}$F spins, and by
the introduction of two new techniques:
an efficient and effective temporal labeling scheme 
for initial state preparation, and a method for precise simultaneous 
rotations of multiple spins at nearby frequencies~\cite{steffen}.

The order of a permutation $\pi$ can be understood via the following analogy:
imagine $2^n$ rooms and $2^n$ {\it one-way} passages connecting 
the rooms, with {\it exactly} one entrance and one exit in each room (for 
some rooms, the passage going out may loop back to the room itself). These 
rules ensure that when making transitions from one room to the next going 
through the passages, you must eventually come back to the room you started 
from. Define the order $r$ as the {\it minimum} number of 
transitions needed to return to the starting room $y$, where $r$ may 
depend on $y$. The order-finding problem is to determine $r$ solely by trials 
of the type ``make $x$ transitions using $\pi$ starting from room $y$ and 
check which room you are in''. Mathematically, we will describe such
trials as queries of an oracle or black box which outputs $\pi^x(y)$. 
The goal then is to find $r$ with the least possible number of queries. 

Cleve showed that order-finding using no other resource or information than 
the oracle is hard both for deterministic 
and probabilistic classical computers~\cite{Cleve99}, i.e. there exists a 
lower bound on the number of oracle queries needed for order-finding which 
is exponential in $n$.
In contrast, this problem can be solved much faster on a quantum computer, 
because finding the order of $\pi(y)$ is equivalent to 
finding the period of the function $f(x) = \pi^x(y)$. The latter can be done 
with a constant probability of success in a constant number of function 
evaluations using a generalization of Shor's quantum 
algorithm~\cite{Kitaev95}.
This is because, in some sense, the quantum computer can make transitions to 
many rooms at once.
Thus, in terms of the number of oracle queries required, the gap between
the quantum and classical case is exponential~\cite{Cleve99}.

We experimentally implemented the order-finding quantum algorithm to
determine the order
of a representative subset of all $4!=24$ permutations on $4$ elements,
including instances of each possible order. It can be proven that the best
classical algorithm needs two queries of the oracle to determine $r$
with certainty, and that using only one query of the oracle, the
probability of finding $r$ using a classical algorithm can be no more than
$1/2$. One optimal classical
strategy is to first ask the oracle for the value of $\pi^3(y)$: when
the result is $y$, $r$ must be $1$ or $3$; otherwise $r$ must be $2$ or $4$.
In either case, the actual order can be guessed only with
probability $1/2$. In contrast, the probability of success is $\sim 0.55$ with
only one oracle-query using the quantum algorithm on a single quantum
computer.
In fact, since in our implementation an ensemble of $\sim 10^{18}$ quantum computers
contribute to the signal, our output data enables the
order to be deduced with virtual certainty. 

\begin{figure}[t]
\noindent \mbox{\psfig{file=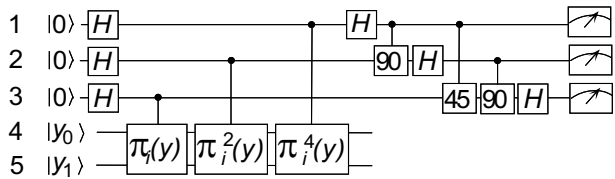,width=3.2in}}
\vspace*{2ex}
\caption{\narrowtext The quantum circuit for order-finding.
Horizontal lines represent qubits; time goes from left to right.
The boxed $90$ and $45$ represent rotations about $\hat{z}$ over those angles.
A black dot connected to a box on another horizontal line indicates that 
the boxed operation is executed if and only if the qubit indicated by
the black dot is $\ket{1}$.
The transformation $\ket{x}\ket{y} \mapsto \ket{x}\ket{\pi^x(y)}$ (step 
{\bf 2}) was implemented in three steps, using the fact that $\pi^x = 
\pi^{x_0} \pi^{2 x_1} \pi^{4 x_2}$, where $x_2x_1x_0$ is the binary 
representation of $x$. Each of these three operations is a permutation on 
qubits $4$ and $5$, controlled by qubits $3$, $2$ and $1$ respectively. The 
details of the controlled permutations depend on $\pi$.
The QFT (step {\bf 3}) was implemented using the construction 
of~\protect\cite{coppersmith}, which swaps the output state of 
qubits $1$ and $3$ compared to the definition of the QFT.}
\label{fig:circuit}
\end{figure}

\begin{figure}[htbp]
\begin{center}
\mbox{\psfig{file=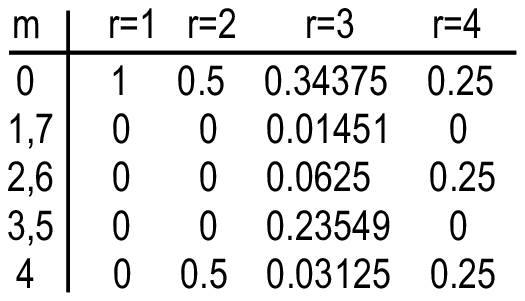,width=1.5in}} \hspace*{3ex}~
\raisebox{1ex}{\psfig{file=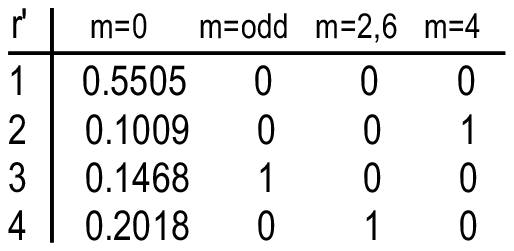,width=1.5in}}
\end{center}
\vspace*{-1ex}
\caption{\narrowtext (Left) The probabilities that the measurement
result $m$ is $0, 1, \ldots,$ or $7$, given $r$ (for an ideal single
quantum computer). (Right) The optimal probabilities with which to
make a guess $r'$ for $r$, given $m$.} 
\label{tab:outcomes}
\end{figure}

The quantum algorithm is as follows (see Fig.~\ref{fig:circuit} 
for the quantum circuit): 
{\bf (0)} initialize the first register of three qubits in 
the state $\ket{0}$ and set the second register of two qubits to 
$\ket{y_1 y_0}$ or for short $\ket{y}$, where 
$y_1 y_0$ is the binary representation of the number $y$;
{\bf (1)} apply a Hadamard transform $H$ to qubits $1$, $2$ and $3$, which puts
the first register in the state $\ket{x} = (\ket{0} + \ket{1} + \ldots +
\ket{7}) / \sqrt{8}$;
{\bf (2)} apply the unitary transformation $\ket{x}\ket{y} \mapsto 
\ket{x}\ket{\pi^x(y)}$, which is the oracle query;
{\bf (3)} perform the quantum Fourier transform (QFT) on the first three
 qubits~\cite{Weinstein99};
{\bf (4)} measure the first three qubits --- for an ideal single quantum 
computer, the possible measurement outcomes 
$m$ and their probabilities are listed in Fig.~\ref{tab:outcomes}
for each possible value of $r$ (the translation to ensemble 
averaged measurements is discussed later);
{\bf (5)} depending on the 
measurement outcome, make a probabilistic guess $r'$ as shown also in 
Fig.~\ref{tab:outcomes}. It is easy to verify that Pr[$r'=r$] is 
$\sim 0.55$, regardless of the probability distribution of $r$ or $\pi$. 

\vspace*{-3ex}

\begin{figure}[h]
\begin{center}
\mbox{\psfig{file=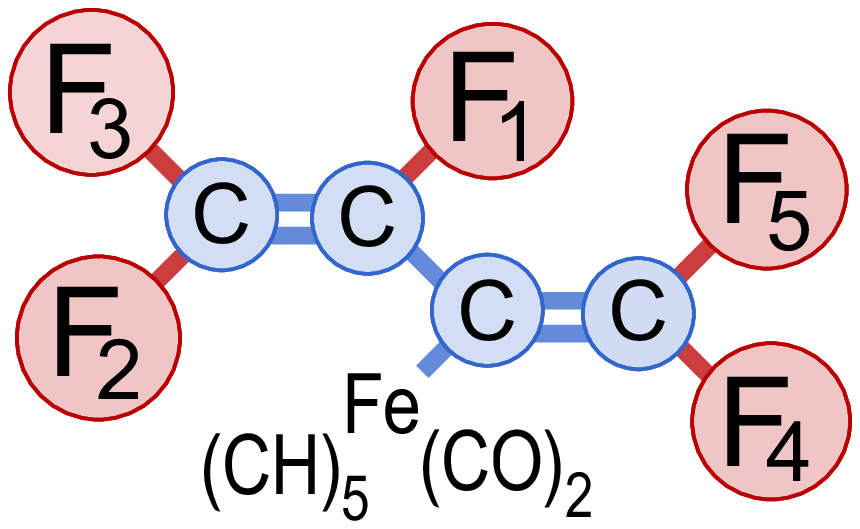,width=1.1in}} \hspace*{4ex}
\mbox{\psfig{file=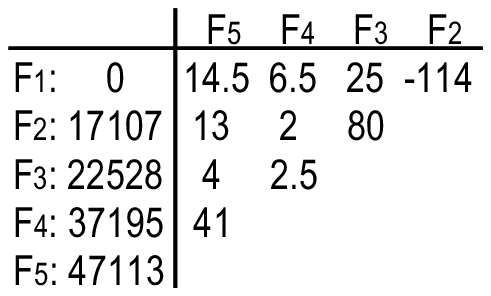,width=1.4in}}
\end{center}
\vspace*{-2ex}
\caption{Structure of the pentafluorobutadienyl 
cyclopentadienyldicarbonyliron complex, with a table of the relative 
chemical shifts of the $^{19}$F spins at $11.7$ T [Hz], and the $J$-couplings 
[Hz]. A total of $76$ out of the $80$ lines in the $5$ spectra are resolved.}
\label{fig:molecule}
\end{figure}

In order to implement this algorithm, we custom synthesized a
molecule~\cite{Green68} containing five $^{19}$F spins which served as the
qubits (Fig.~\ref{fig:molecule}). When placed in a static magnetic field,
each spin has two discrete energy eigenstates, spin-up, $\ket{0}$, and
spin-down, $\ket{1}$, described by the Hamiltonian $\hbar \omega_i I_{zi}$,
where $\omega_i$ is the transition frequency between the spin-up and
spin-down states and $I_z$ is the $\hat{z}$ component of the spin angular
momentum operator. In this molecule, {\em all} five spins are remarkably
well-separated in frequency, $\omega_i$, and are mutually coupled
with a coupling Hamiltonian of the form $2 \pi\hbar J_{ij} I_{zi} I_{zj}$
(Fig.~\ref{fig:molecule}). The linewidths of the NMR transitions are
$\sim$ 1 Hz, so the $T_2$ quantum coherence times of the spins were at least
$\sim$ 0.3~s. The $T_1$ time constants were measured to be between $3$ and
$12$ s.

\vspace*{-1ex}

\vspace*{-2ex}
This nuclear spin system was used at room temperature. The thermal 
equilibrium state is then highly mixed, 
i.e. the probabilities that each spin is $\ket{0}$ or
$\ket{1}$ differ by only $1$ part in $\sim 10^5$, which is not a suitable
initial state for a quantum computation~\cite{initial_state}. 
Instead, as has previously been
shown, an ``effective pure'' initial state must be created, in which only
the spins in the $\ket{00000}$ state produce a net output
signal~\cite{Gershenfeld97,Cory97}. We devised a new procedure to prepare
effective pure states which is best understood in terms of product
operators~\cite{freeman}. The equilibrium density matrix for a homonuclear
spin system is a sum of $n=5$ terms: $IIIIZ + IIIZI + IIZII + IZIII +
ZIIII$. The desired effective pure state density matrix is $ IIIIZ + \ldots
+ ZIIII + IIIZZ + \ldots + ZZIII + IIZZZ + \ldots + ZZZII + IZZZZ + \ldots +
ZZZZI + ZZZZZ$, a sum of $2^n-1=31$ terms. Using short sequences of
controlled-NOT operations (${\sc C}_{ij}$ flips spin $j$ if and only if $i$
is $\ket{1}$), the five terms obtained in
equilibrium can be transformed into different sets of five terms. For
homonuclear spin systems, the
summation of only $\lceil (2^n-1)/n \rceil = 7$ different experiments 
thus suffices to create all $31$ terms, although it may be advantageous 
to use slightly more experiments in order to keep the preparation sequences 
short. 
In contrast, both conventional temporal averaging~\cite{temp-lab} and later 
improvements~\cite{Vandersypen00} require up to $2^n-1$ experiments, and 
furthermore suffer from higher complexity and/or a lower signal-to-noise 
ratio. 
The overhead is still exponential though, so even this improved technique is 
not scalable. But importantly, a scalable approach to NMR quantum computation
exists~\cite{Schulman99} and may become practical if large
polarization enhancements can be achieved.
We used $9$ experiments, giving a total of $45$ product operator terms. 
The $14$ extra terms were canceled out pairwise, using {\sc NOT} 
(${\sc N}_i$) operations to flip the sign of selected terms. The $9$ state 
preparation sequences were

$
{\sc C}_{51} {\sc C}_{45} {\sc C}_{24} {\sc N}_3, 
{\sc C}_{14} {\sc C}_{31} {\sc C}_{53} {\sc N}_2, 
{\sc C}_{54} {\sc C}_{51} {\sc N}_2, \\
\indent {\sc C}_{31} {\sc C}_{43} {\sc C}_{23} {\sc N}_5, 
{\sc C}_{21} {\sc C}_{52} {\sc C}_{45} {\sc C}_{34}, 
{\sc C}_{53} {\sc C}_{25} {\sc C}_{12} {\sc N}_4, \\
\indent {\sc C}_{12} {\sc C}_{15} {\sc C}_{13} {\sc C}_{41}, 
{\sc C}_{32} {\sc C}_{13} {\sc C}_{25} {\sc N}_4,
{\sc C}_{35} {\sc C}_{23} {\sc N}_1.
$

\begin{figure}[h]
\noindent \mbox{\psfig{file=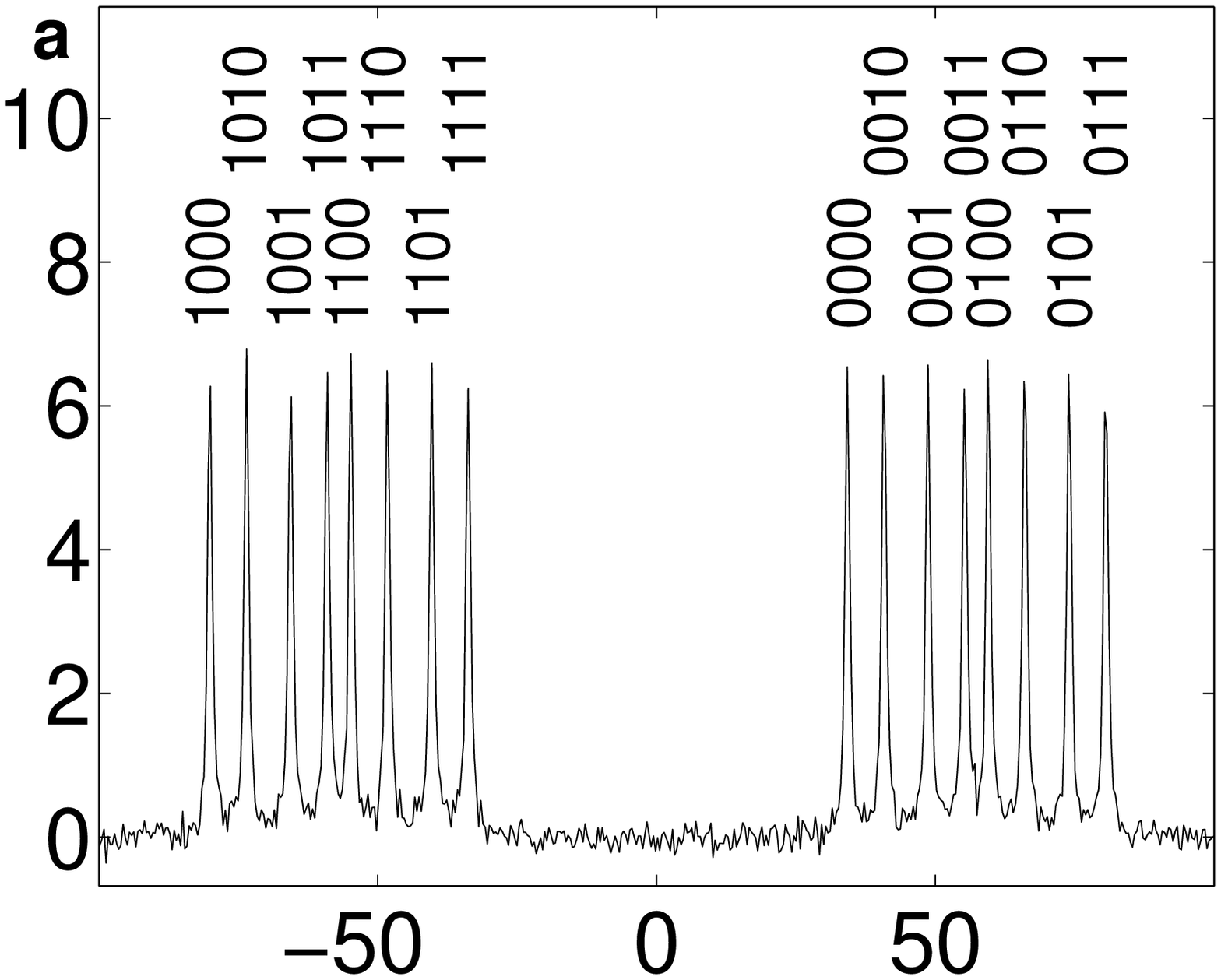,height=1.4in}}
\mbox{\psfig{file=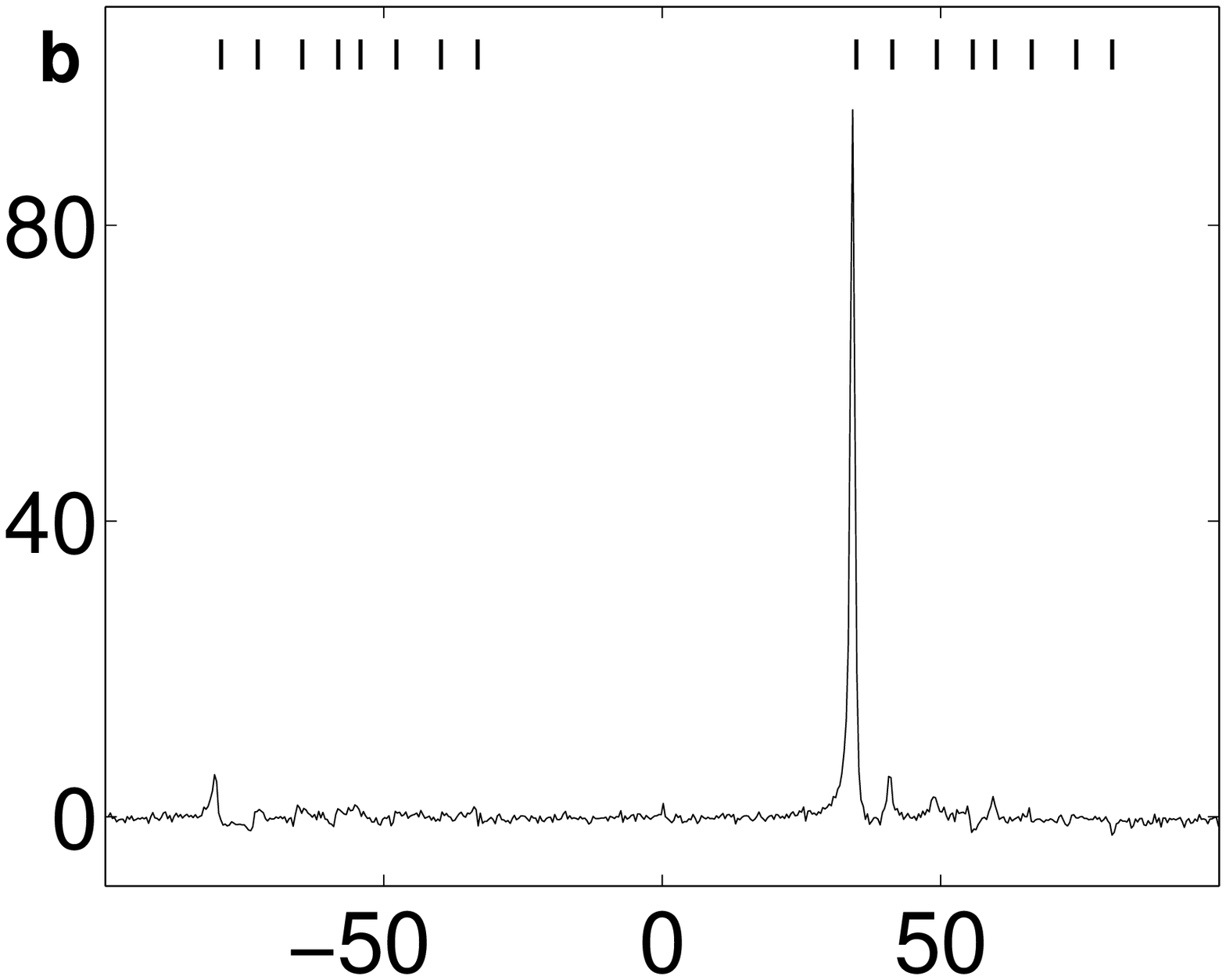,height=1.4in}}
\vspace*{1ex}
\caption{All spectra shown here and in Fig.~\protect\ref{fig:spectra}
display the real part of the spectrum in the same arbitrary units, and were
obtained without phase cycling or signal-averaging (except for
Fig.~\protect\ref{fig:spectra} c, where 16 identical scans were averaged). A
$0.1$ Hz filter was applied. Frequencies are in units of Hz with respect to
$\omega_1$. {\bf a}, the spectrum of spin $1$ in equilibrium. The $16$ lines
are due to shifts in the transition frequency $\omega_1$ by $\pm J_{1j}/2$,
depending on whether spin $j$ is in $\ket{0}$ or $\ket{1}$. In equilibrium,
all the $32$ states are nearly equally populated, hence the $16$ lines in
each spectrum have virtually the same intensity. Taking into account the
sign and magnitude of the $J_{1,j}$, the $16$ lines in the spectrum of spin
$1$ can be labeled as shown. {\bf b}, the same spectrum when
the spins are in an effective pure state. Only the line labeled $0000$ is
present.}
\label{fig:labeling}
\end{figure}

\noindent The resulting data are remarkably clean, as illustrated in 
Fig.~\ref{fig:labeling}. After the state preparation, only the $0000$ line 
should remain visible, reflecting that only molecules with all spins in the 
ground state contribute to the signal. This is clearly observed in the 
measured spectra.

\vspace*{-2ex}

\vspace*{-2ex}
This effective pure state served as the initial state for the
order-finding algorithm. The actual computation was realized via a sequence
of $\sim 50$ to $\sim 200$ radio-frequency (RF) pulses, separated by time
intervals of free evolution under the Hamiltonian, for a total duration
of $\sim 50$ to $\sim 500$ ms, depending on $\pi$. 
The pulse sequences for the order-finding algorithm were designed by
translating the quantum circuits of Fig.~\ref{fig:circuit} into one-
and two-qubit operations, employing several simplification 
methods~\cite{Vandersypen00}.
These pulse sequences were implemented on a custom 
modified four-channel Varian Unity INOVA spectrometer, and a Nalorac HFX 
probe. The frequency of one channel was set at $(\omega_2 + \omega_3)/2$, and 
the other three channels were set on the resonance of spins $1, 4$ and $5$. 
The chemical shift evolutions of spins $2$ and $3$ were calculated with
the help of a time-counter, which kept track of the time elapsed from the 
start of the pulse sequence. On-resonance excitation of spins $2$ and $3$ 
was achieved using phase-ramping techniques~\cite{patt}. All pulses
were spin-selective and Hermite shaped~\cite{freeman}. 
Rotations about the $\hat{z}$-axis were implemented by adjusting the 
phases of the subsequent pulses~\cite{freeman}. Unintended 
phase shifts~\cite{emsley} of spins $i$ during a pulse on spin $j \neq i$ 
were calculated and accounted for by adjusting the phase of subsequent 
pulses. During simultaneous pulses, the effect of these phase shifts 
was largely removed by shifting the frequency of the pulses via phase-ramping. 
The pulse frequency shifts are designed such that they track the shifting 
spin frequencies and thereby greatly improve the accuracy of the simultaneous
rotations of two or more spins~\cite{steffen}.
This technique circumvents the need to avoid simultaneous pulses at nearby 
frequencies~\cite{linden}, and thus permits more efficient pulse
sequences.

Upon completion of the pulse sequence, the states of the three spins in the
first register were measured and the order $r$ was determined from the
read-out. Since an ensemble of quantum computers rather than a single
quantum computer was used, the measurement gives the bitwise average values 
of $m_i (i=1,2,3)$, instead of a sample of $m=m_1 m_2 m_3$ with probabilities 
given in Fig.~\ref{tab:outcomes}~\cite{bitwise_average}. 
Formally, measurement of spin $i$ returns $O_i = 1
- 2 \langle m_i \rangle = 2\mbox{Tr}(\rho I_{zi})$, where $\rho$ is the
final density operator of the system. The $O_i$ are obtained experimentally
from integrating the peak areas in the spectrum of the magnetic signal of
spin $i$ after a $90^\circ$ read-out pulse on spin $i$, phased such that
positive spectral lines correspond to positive $O_i$. The theoretically 
predicted
values of $O_i$ ($i=1,2,3$) for each value of $r$ follow directly from the
probabilities for $m$ in Fig.~\ref{tab:outcomes}. For reference, we also
include the values of $O_4$ and $O_5$ (for $y=0$; if $y \neq 0$, $O_4$ and
$O_5$ can be negative): for the case $r=1$ the $O_i$ are $1, 1, 1, 1, 1$;
for $r=2$ they are $1, 1, 0, 1, 0$; and for $r=4$ they are $1, 0, 0, 0, 0$.
For $r=3$, the $O_{i}$ ($i=1,2,3$) are $0, 1/4, 5/16$, and $O_{4}$ and $O_5$
can be $0$, $\pm 1/4$ or $\pm 1/2$, depending on $y$. The value of $r$ can
thus be unambiguously determined from the spectra of the three spins in the
first register. This was confirmed experimentally by taking spectra for these
three spins, which were in excellent agreement with the theoretical
expectations.

\begin{figure}[h]
\begin{center}
\hspace*{-2ex} $ \noindent \begin{array}{ll}
\noindent \mbox{\psfig{file=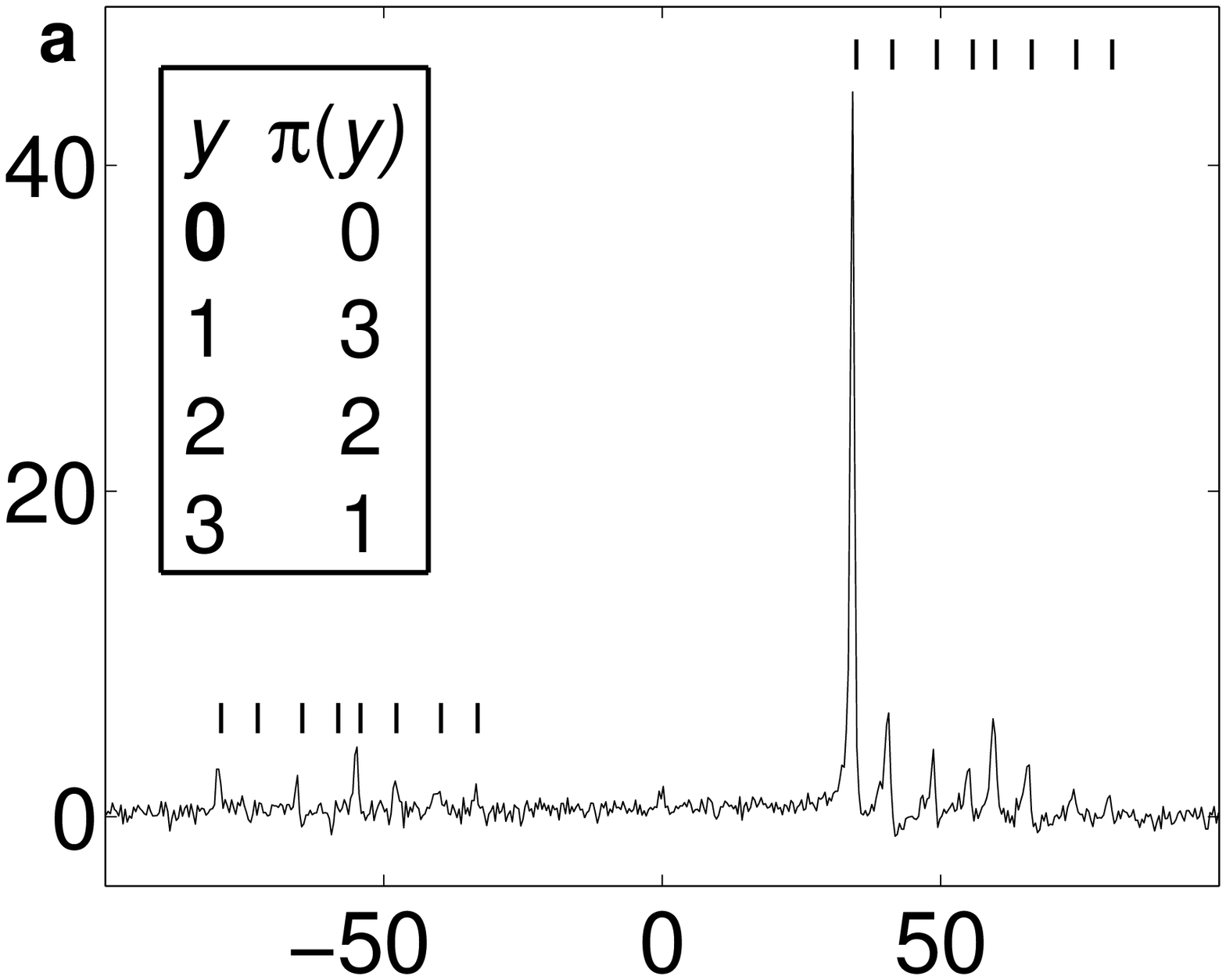,height=1.4in}} &
\mbox{\psfig{file=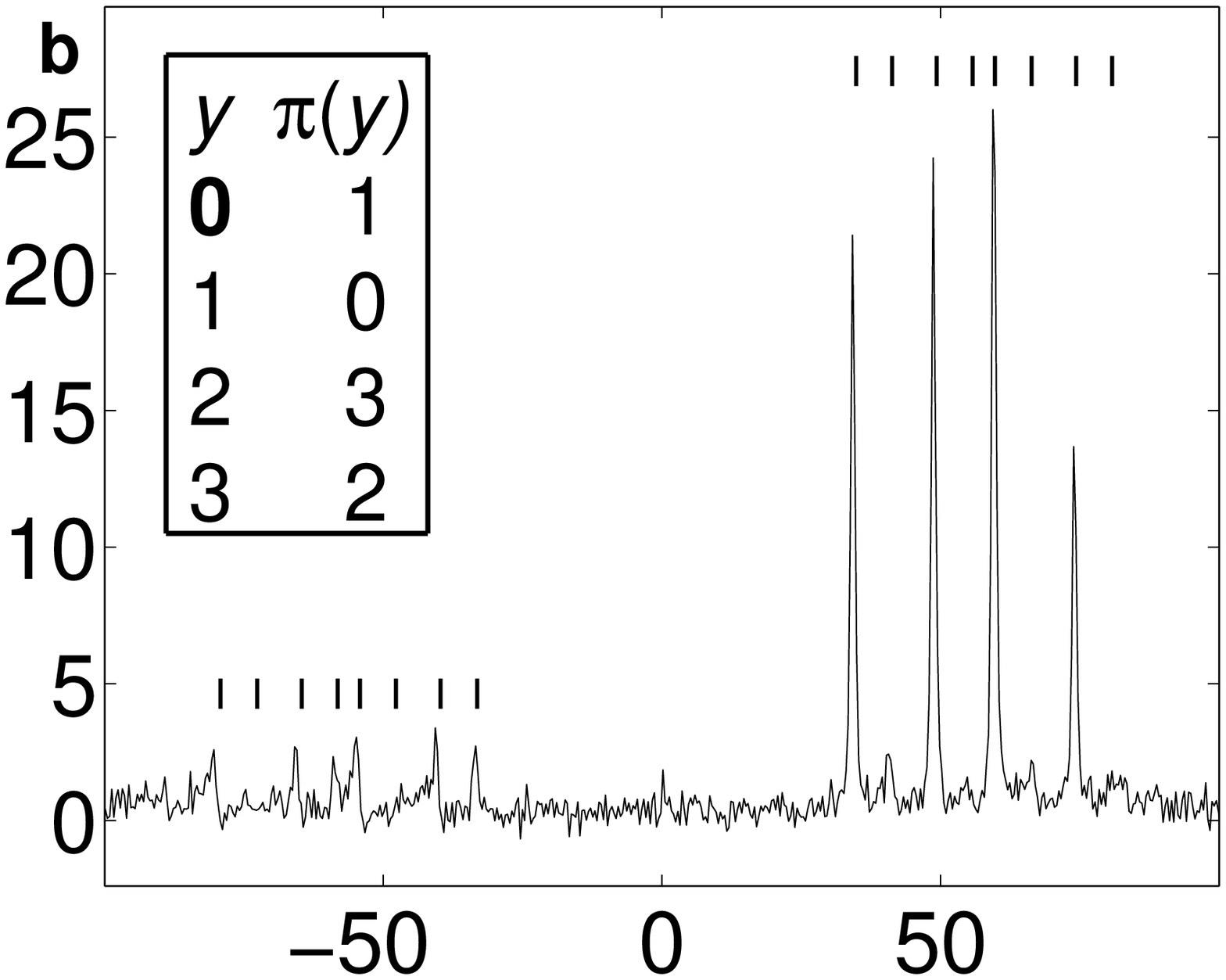,height=1.4in}} \vspace*{-1ex} \\
\noindent \mbox{\psfig{file=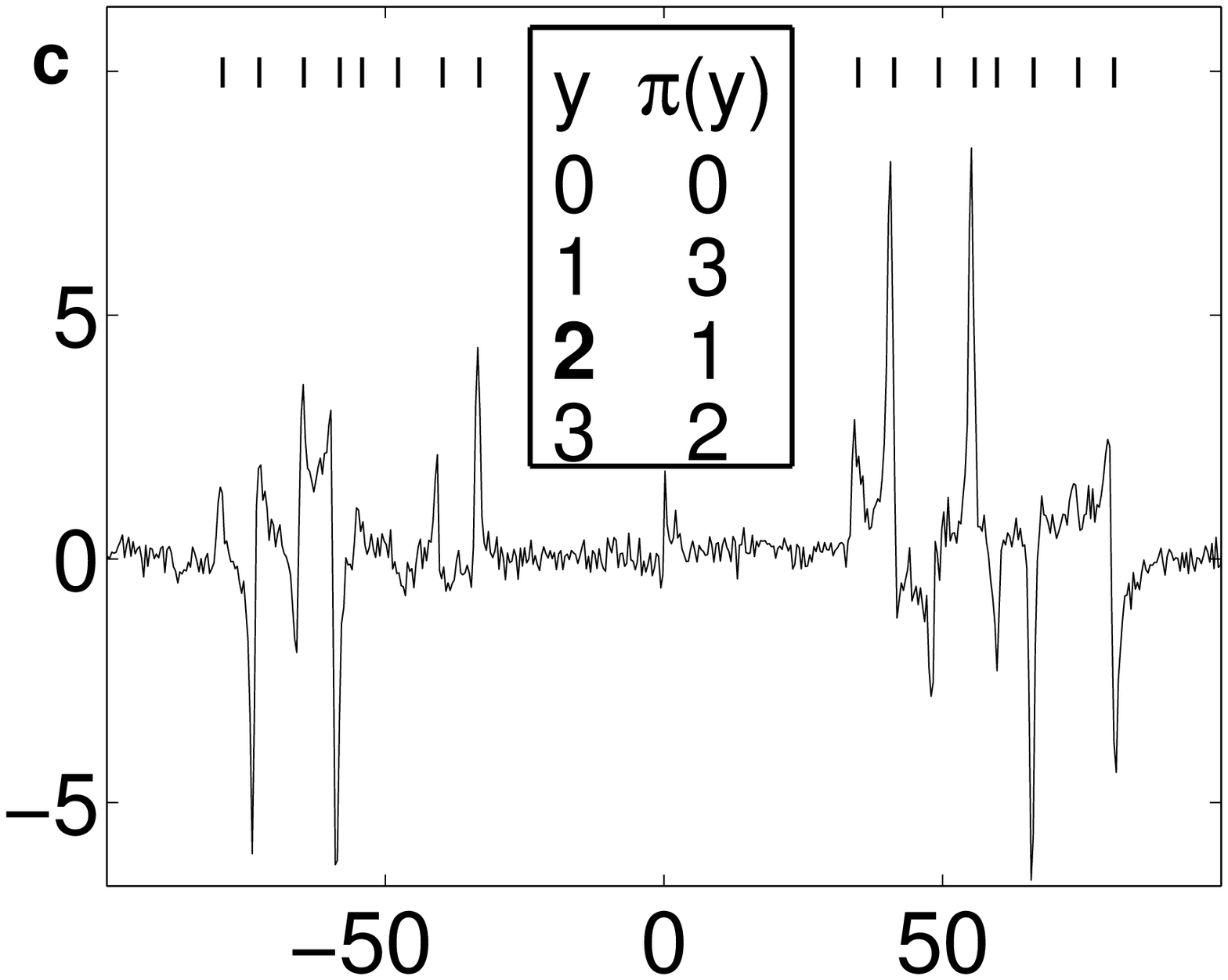,height=1.4in}} &
\mbox{\psfig{file=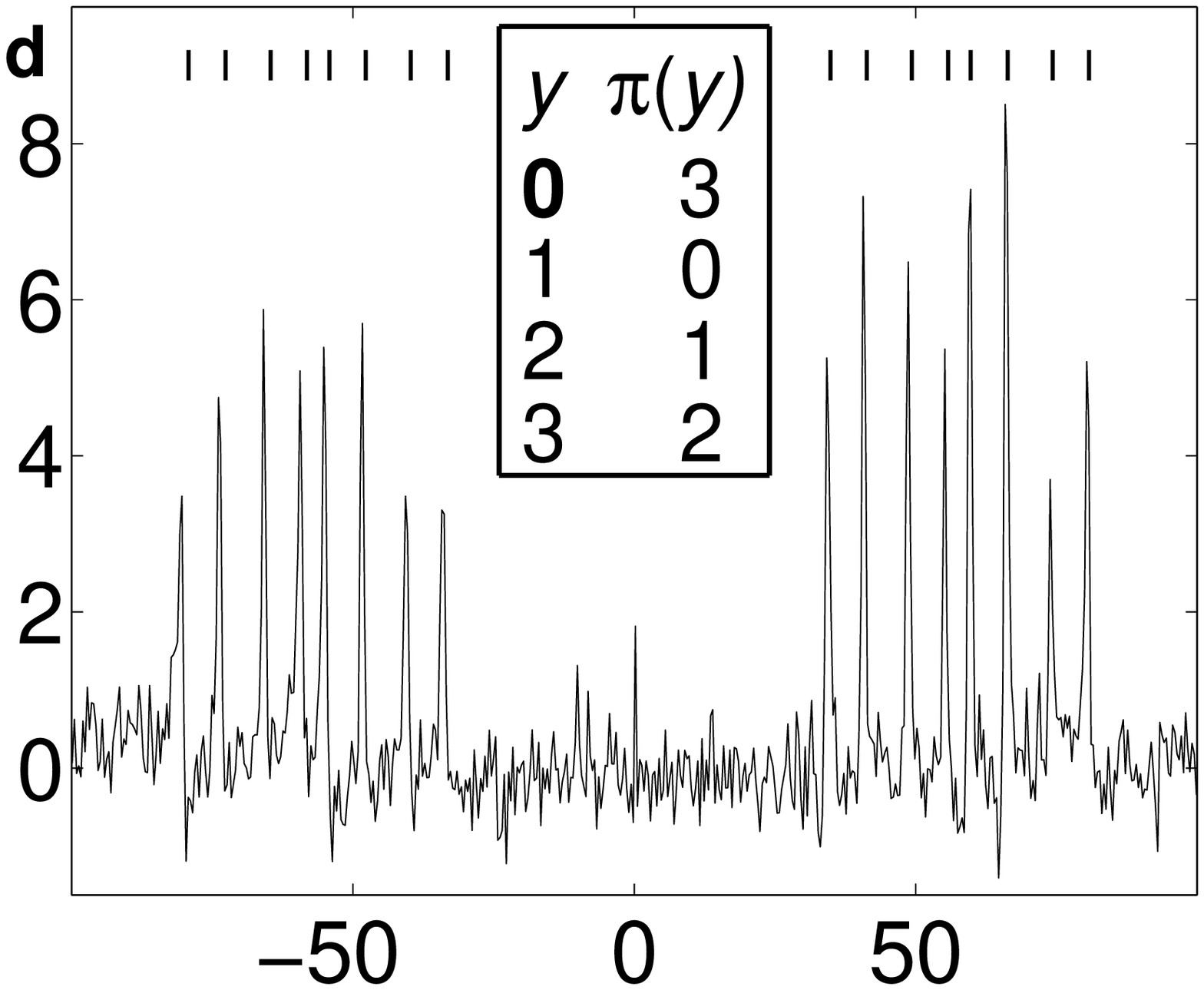,height=1.4in}}
\end{array} $
\end{center}
\vspace*{-2.5ex}
\caption{Spectra of spin $1$ acquired after executing the order-finding 
algorithm. The respective permutations are shown in inset, with the input
element highlighted.
The $16$ marks on top of each spectrum indicate the position of the $16$ lines 
in the thermal equilibrium spectrum.
The transformation $\ket{x}\ket{y} \mapsto \ket{x}\ket{\pi^x(y)}$ is 
realized by
{\bf a}, $r=1$: P$_{54}$ C$_{35}$ P$_{54}^\dagger$ C$_{35}$ P$_{34}$ (P$_{ij}$ rotates spin $j$ by $90^\circ$ about $\hat{z}$ if and only if spin $i$ is $\ket{1}$).
{\bf b}, $r=2$: C$_{35}$.
{\bf c}, $r=3$: C$_{32}$ C$_{25}$ C$_{32}$ C$_{21}$ P$_{14}$ C$_{51}$ P$_{14}^\dagger$ C$_{51}$ P$_{54}$ C$_{21}$ P$_{15}$ C$_{41}$ P$_{15}^\dagger$ C$_{41}$ P$_{45}$. (this sequence does the transformation $\pi^x(y)$ for $y=2$ only;
sequences for $r=3$ that would work for any $y$ are 
prohibitively long). 
{\bf d}, $r=4$: C$_{24}$ P$_{34}$ P$_{54}$ C$_{35}$ P$_{54}$. 
Each transformation was tested independently to confirm its proper operation.}
\label{fig:spectra}
\end{figure}

In fact, the complete spectrum of any one of the first three spins uniquely 
characterizes $r$ by virtue of extra information contained in the splitting 
of the lines. For the spectrum of spin $1$ 
the values of $O_i$ given above indicate that for $r=1$, only the $0000$ 
line (see Fig.~\ref{fig:labeling}) will be 
visible since spins $2-5$ are all in $\ket{0}$. Furthermore, this
line should be absorptive and positive since spin $1$ is also in $\ket{0}$.
Similarly, for $r=2$ the $0000, 0001, 0100$ and $0101$ lines are expected 
to be positive, and for $r=4$ all $16$ 
lines should be positive. Finally, for $r=3$, the {\it net} area under the 
lines of spin $1$ should be zero since $O_1=0$, although most individual 
lines are expected to be non-zero and partly dispersive.
These unambiguous characteristics are reflected in the data. Results for four 
representative permutations are presented in Fig.~\ref{fig:spectra}. 
In all cases, the spectrum is in good agreement with the predictions, both 
in terms of the number of lines, and their position, sign and amplitude. 
Slight deviations from the ideally expected spectra are attributed mostly 
to incomplete refocusing of undesired coupled evolutions and to decoherence. 

\vspace*{-2ex}

\vspace*{-2ex}
The success of the order-finding experiment required the synthesis of a 
molecule with unusual NMR properties and the development of several new 
methods to meet the increasing demands for control over the spin 
dynamics. The major difficulty was to address and control the qubits 
sufficiently well to remove undesired couplings while leaving select 
couplings active. Furthermore, the pulse sequence had to be executed 
within the coherence time. Clearly, the same challenges 
will be faced in moving beyond liquid state NMR, and we anticipate that 
solutions such as those reported here will be useful in future quantum 
computer implementations, in particular in those involving spins, such as 
solid state NMR~\cite{Kane98}, electron spins in quantum dots~\cite{Loss98} 
and ion traps~\cite{Sackett00}.

We thank X. Zhou, A. Verhulst, M. Sherwood, S. Smallcombe, A. Brooke
and R. Laflamme for help and discussions, J. Harris and W. Risk for
support, and the Aspen Center for Physics for its
hospitality. L.V. gratefully acknowledges support by a Yansouni Family
Stanford Graduate Fellowship. This work was supported by DARPA under
the NMRQC initiative.

\vspace*{1ex}
$^*$ Email address: lieven@snow.stanford.edu.

\end{multicols}


\begin{thebibliography}{0}
\bibitem{initial_state} If $r$ were independent of the value of $y$, the 
second register would not have to be prepared in an effective pure state,
but in general $r$ may depend on $y$ so all five spins were made 
effective pure in all spectra shown.
\bibitem{bitwise_average} It is not clear, however, that the bitwise 
average outputs of the QFT suffice to determine $r$ for permutations 
on arbitrary $n$. Instead, classical post-processing on the quantum 
computer~\cite{Gershenfeld97} can be used to compute $r$.
\end{thebibliography}
\end{document}